\documentclass[twocolumn]{jpsj3}
\usepackage{txfonts}
\usepackage{bm}
\usepackage{color}


\title{Emergence of Orbital Nematicity in the Tetragonal Phase of BaFe$_2$(As$_{1-x}$P$_x$)$_2$}

\author{Tetsuya Iye$^{1,2}$\thanks{E-mail: tiye@scphys.kyoto-u.ac.jp}, 
Marc-Henri Julien$^{1}$\thanks{E-mail: marc-henri.julien@lncmi.cnrs.fr}, 
Hadrien Mayaffre$^{1}$, 
Mladen Horvati$\rm\acute{c}$$^{1}$, 
Claude Berthier$^{1}$, 
Kenji Ishida$^{2}$\thanks{E-mail: kishida@scphys.kyoto-u.ac.jp},
Hiroaki Ikeda$^{2}$, 
Shigeru Kasahara$^{2}$, 
Takasada Shibauchi$^{2}$, 
and Yuji Matsuda$^{2}$}

\inst{$^1$Laboratoire National des Champs Magn\'etiques Intenses, LNCMI - CNRS (UPR3228), UJF, UPS and INSA, BP 166, 38042 Grenoble Cedex 9, France \\
       $^2$Department of Physics, Graduate School of Science, Kyoto University, Kyoto 606-8502, Japan} %\\

\abst{
We report on $^{75}$As-NMR measurements in single crystalline BaFe$_{2}$(As$_{0.96}$P$_{0.04}$)$_{2}$ for magnetic fields parallel to the orthorhombic $[110]_{\rm o}$ and $[100]_{\rm o}$ directions above the structural transition temperature $T_{\rm S}\simeq121$ K. 
A large difference in the linewidth between the two field directions reveals in-plane anisotropy of the electric field gradient, even in the tetragonal phase. 
This provides microscopic evidence of population imbalance between As-$4p_{x}$ and $4p_{y}$ orbitals, which reaches $|n_x-n_y|/|n_x+n_y| \sim 15$\% at $T\rightarrow T_{\rm S}$ and is a natural consequence of the orbital ordering of Fe-3$d_{xz}$ and $d_{yz}$ electrons. 
Surprisingly, this orbital polarization is found to be already static near room temperature, suggesting that it arises from the pinning of anisotropic orbital fluctuations by disorder. 
The effect is found to be stronger below $\sim 160$ K, which coincides with the appearance of nematicity in previous torque and photoemission measurements. 
These results impose strong constraints on microscopic models of the nematic state.}

\begin{document}
\maketitle
%\section{Introduction}
The relationships among spin, orbital, and lattice degrees of freedom are a central issue in iron pnictides since fluctuations in these channels are potentially at the origin of high-$T_{\rm c}$ superconductivity. 
The problem is best illustrated by the presence of two interrelated structural and magnetic transitions, at temperatures $T_{\rm S}$ and $T_{\rm N}$, respectively, which both disappear near the optimal $T_{\rm c}$ in most pnictides~\cite{HashimotoSCIENCE336,NingPRL,YoshizawaJPSJ,NakaiPRL}. 
Although the tetragonal-to-orthorhombic transition slightly precedes (or coincides with) the antiferromagnetic (AFM) transition, the latter is not necessarily a consequence of the former: whether the orthorhombic distortion is driven by magnetic or orbital fluctuations has been the subject of continuing debate\cite{FernandesSST} (see Ref. 5 and references therein). 
Furthermore, the tetragonal phase above $T_{\rm S}$ turns out to be unconventional: various electronic properties have been found to break the fourfold symmetry of the Fe-As planes ~\cite{ChuSCIENCE,ChuSCIENCE337,TanatarPRB,BlombergPRB,BlombergNATURECOM,JeschePRB86,DuszaEPL,NakajimaPRL,YiPNAS,RosenthalARXIV,KimPRL,HarrigerPRB,LuoPRL,JiangPRL,GallaisARXIV,YangARXIV,KasaharaNATURE486,KWSongPRB,ShimojimaARXIV,SinghARXIV}. 
The relative role and the microscopic origin of the spin and orbital instabilities leading to such an ``electronic nematic state" are also controversial~\cite{FangPRB77,XuPRB78,FernandesPRB85,KWSongPRB,LeePRL,ChenPRB,LvPRB,KontaniPRB,KrugerPRB,LaadPRB,StanevPRB,WysockiNATUREPHYS,YinNATUREPHYS}.
Actually, because magnetic and orbital degrees of freedom are entangled and most likely cooperate~\cite{LiangPRL}, it is challenging to determine which one, if any, is dominant. 
While both the magnetic and orbital scenarios find support in experiments ~\cite{FernandesSST,FernandesPRL107,FernandesPRL111,KontaniSSC,KontaniARXIV,YiPNAS,ChuSCIENCE,YoshizawaJPSJ,BohmerARXIV,GallaisARXIV}, the nematic state remains puzzling.

Here, we report on NMR measurements above $T_{\rm S}$ in underdoped BaFe$_{2}$(As$_{0.96}$P$_{0.04}$)$_{2}$. 
We demonstrate that orbital polarization of the As-4$p$ orbitals, related to Fe-3$d$ polarization, is present within electronic domains in the tetragonal phase, even without any applied uniaxial stress. 
We estimate the magnitude of this polarization and show that it is static. 
We further reveal an unanticipated temperature dependence: while the onset of the nematic state, as inferred from torque and photoemission experiments~\cite{KasaharaNATURE486,ShimojimaARXIV}, is manifested by an upturn near 160~K in our linewidth data, static orbital polarization is found to be already present at much higher temperatures. 
This suggests that static short-range orbital order is first nucleated around defects and that its evolution towards longer-range order is directly involved in the manifestation of nematicity at the macroscopic scale.

\begin{figure}[t]
\begin{center}
\includegraphics[width=\hsize,clip]{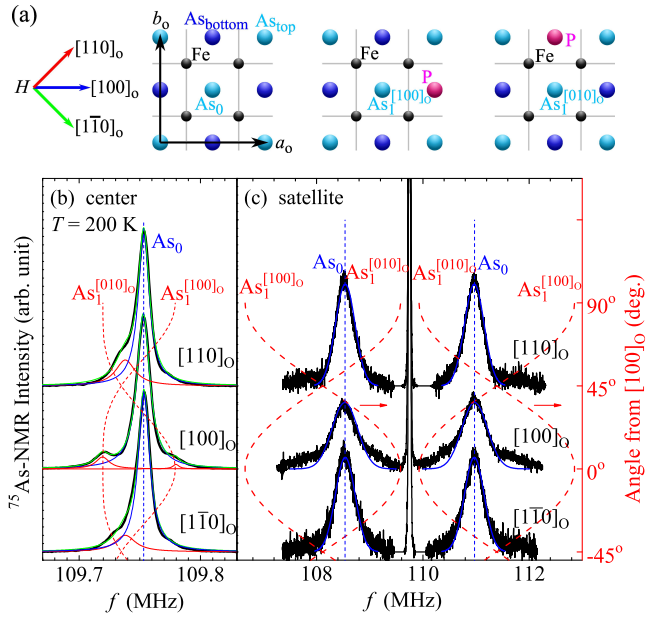}
\caption{(Color online) (a) Fe$Pn$ layers and applied field directions. Left panel: the central As site without P dopant among its NN is named As$_{0}$. Middle and right panels: the central As sites having one P dopant among their NN in the $[100]_{\rm o}$ and $[010]_{\rm o}$ directions are named As$_{1}^{[100]_{\rm o}}$ and As$_{1}^{[010]_{\rm o}}$, respectively. $^{75}$As-NMR central line (b) and satellites (c) at 200~K for different field orientations. Continuous lines are fits to As$_{0}$ (blue) and As$_{1}$ (red) sites. Expected resonance positions are shown as dashed lines.
}
\label{fig:AsRotationSP}
\end{center}
\end{figure}

%\section{Experimental Methods}
Single-crystalline BaFe$_{2}$(As$_{0.96}$P$_{0.04}$)$_{2}$ was synthesized by the conventional self-flux method\cite{KasaharaPRB}.
Edges of the crystal were cut along $[110]_{\rm o}$ in the orthorhombic notation (As-Fe direction), $[100]_{\rm o}$ (Fe-Fe direction), and $[1\bar{1}0]_{\rm o}$ [see Fig.~\ref{fig:AsRotationSP}(a)] and the resulting dimensions of the crystal were $2.0\times1.4\times0.25$~mm$^{3}$.
For the following NMR measurements, the field was parallel to the $ab$-plane to within $\pm1^{\circ}$. 
$^{75}$As-NMR spectra were obtained by sweeping the frequency in a fixed field of 15~and 6~T.
The structural transition at $T_{\rm S}\simeq 121$~K was determined from a kink in the resistivity measured on a crystal from the same batch while the magnetic ordering at $T_{\rm N}\simeq 121$~K was determined from the sharp drop in the NMR signal intensity, signifying the simultaneous structural and magnetic transition $T_{\rm S}=T_{\rm N}$.

%\section{Results}
%%%%%%%%%%%%%%%%%%%%%%%%%%%%%%%%%%%%%%% Fig.1 %%%%%%%%%%%%%%%%%%%%%%%%%%%%%%%%%%%%%%%%%%%%
%%%%%%%%%%%%%%%%%%%%%%%%%%%%%%%%%%%%%%%%%%%%%%%%%%%%%%%%%%%%%%%%%%%%%%%%%%%%%%%%%%
%$^{75}$As-NMR spectra are measured at 200~K in a field of 15 T along $[110]_{\rm o}$, $[100]_{\rm o}$, and $[1\bar{1}0]_{\rm o}$.

Since $^{75}$As nuclei have spin $I\!=\!3/2$, the spectrum shows three lines corresponding to the transitions $I_{z}=m\leftrightarrow m-1 \ (m=\pm1/2, 3/2)$ at the frequencies $f_{m\leftrightarrow m-1}(\theta,\phi)=(\gamma/2\pi)\mu_{0}H[1+K(\theta,\phi)]+(m-1/2)\nu(\theta,\phi)+\text{(\rm 2$^{\rm nd}$-order quadrupolar correction)}$, where $(\theta,\phi)$ indicates polar and azimuthal angles of the applied field in the orthorhombic basis $(a_{\rm o},b_{\rm o},c)$, $\gamma/2\pi=7.2919$~MHz/T is the $^{75}$As gyromagnetic ratio, and $K(\theta,\phi)$ represents the Knight shift, which is proportional to the magnetic susceptibility $\chi(q\!=\!0,\omega\!=\!0)$.
%%%%%%%%%%%%%%%%%%%%%%%%%%%%%%%%%%%%%%% Fig.2 %%%%%%%%%%%%%%%%%%%%%%%%%%%%%%%%%%%%%%%%%%%%
\begin{figure}[t]
\begin{center}
\includegraphics[width=\hsize,clip]{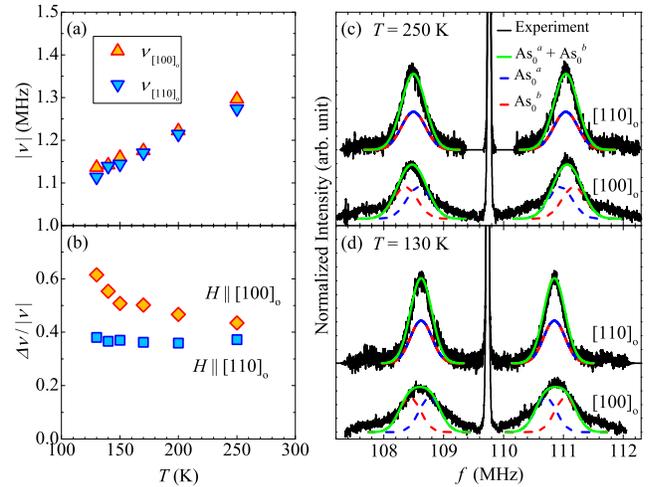}
\caption{(Color online) (a) $T$ dependence of the quadrupole frequency $|\nu|$. (b) $T$ dependence of the distribution ${\Delta\nu}/|\nu|$ where $\Delta\nu$ is the full-width at half-maximum of a satellite line. Fitting error bars are within the symbols. 
$^{75}$As satellites at 250~ (c) and 130~K (d) for $H\parallel[100]_{\rm o}$ and $[110]_{\rm o}$, respectively. Broken blue and red lines represent As$_{0}^{a}$ and As$_{0}^{b}$ sites, respectively, obtained from Gaussian fits to the lineshape (their sum is shown as a solid green line). The two sites are identical for $H\parallel[110]_{\rm o}$, while they split for $H\parallel[100]_{\rm o}$.
}
\label{fig:TemperatureSP}
\end{center}
\end{figure}
%%%%%%%%%%%%%%%%%%%%%%%%%%%%%%%%%%%%%%%%%%%%%%%%%%%%%%%%%%%%%%%%%%%%%%%%%%%%%%%%%%%%%%%%%%
%%%%%%%%%%%%%%%%%%%%%%%%%%%%%%%%%%%%%%% Fig.3 %%%%%%%%%%%%%%%%%%%%%%%%%%%%%%%%%%%%%%%%%%%%
\begin{figure}[t]
\begin{center}
\includegraphics[width=\hsize,clip]{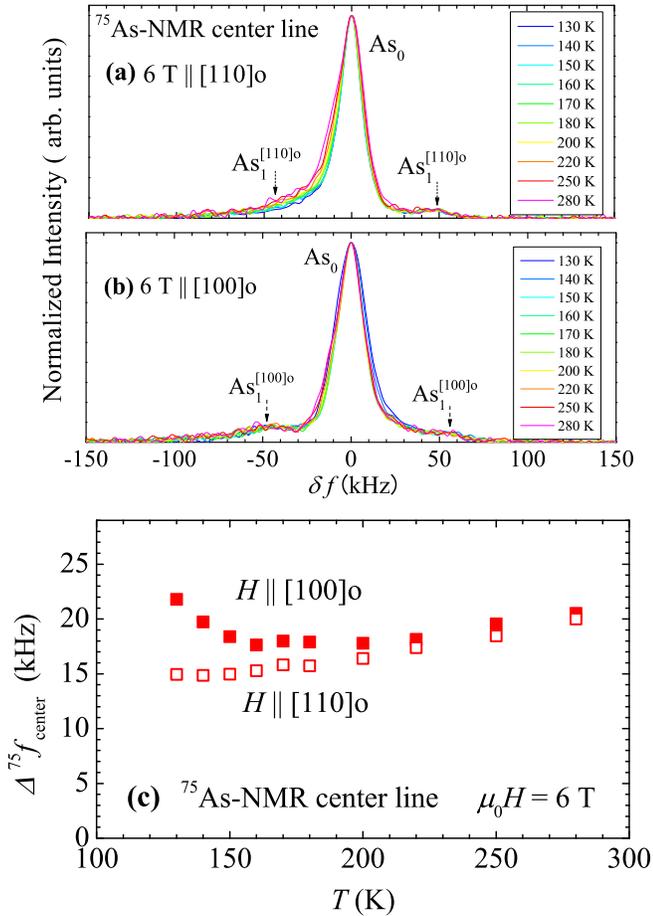}
\caption{(Color online) $T$ dependence of the $^{75}$As-NMR arising from the central transition obtained for $H\parallel[110]_{\rm o}$ (a) and $[100]_{\rm o}$(b). (c) $T$ dependence of the full-width at half-maximum of the centerline from the As$_0$ site in $H\parallel[110]_{\rm o}$ and $[110]_{\rm o}$. $\Delta f_{\rm center}$ for $H\parallel[100]_{\rm o}$ starts to increase below 160 K, as observed in the linewidth of the satellite peak shown in Fig.\;2\,(b).  
}
\label{fig:Fig3C}
\end{center}
\end{figure}
%%%%%%%%%%%%%%%%%%%%%%%%%%%%%%%%%%%%%%%%%%%%%%%%%%%%%%%%%%%%%%%%%%%%%%%%%%%%%%%%%%%%%%%%%%
The angular dependence of the quadrupole frequency $\nu_{\alpha}$ is expressed as $\nu(\theta,\phi)=(\nu_{c}/2)(m-1/2)(3\cos^{2}\theta-1-\eta\sin^{2}\theta\cos2\phi)$. $\nu_{\alpha}$ is proportional to the diagonal components of the electric field gradient (EFG) tensor $V_{\alpha\alpha}\equiv\partial^{2}V/\partial r_{\alpha}^{2}$: $\nu_{\alpha}=eV_{\alpha\alpha}Q/2h \ (\alpha=a,b,c)$, where $Q$ is the electric quadrupole moment.
The EFG tensor of the so-called As$_{0}$ sites far from any P dopant [see Fig.~\ref{fig:AsRotationSP}(a)] is diagonalized in the orthrhombic-basis $(a_{\rm o},b_{\rm o},c)$~\cite{KitagawaJPSJ77}.
The in-plane EFG anisotropy is defined as $\eta\equiv|V_{aa}-V_{bb}|/|V_{cc}|=|\nu_{a_{\rm o}}-\nu_{b_{\rm o}}|/|\nu_{c}|$ with $\nu_{a}+\nu_{b}+\nu_{c}=0$.
Therefore, $\eta=0$ in a local tetragonal environment.

Small additional peaks observed in the tail of the central line [Fig.~\ref{fig:AsRotationSP}(b)] are attributed to those As sites (hereafter called As$_1$) having one P dopant among their four nearest neighbors (NNs). The As$_1$ resonance splits into two peaks for $H\parallel[100]_{\rm o}$ because As$_{1}^{[100]_{\rm o}}$ and As$_{1}^{[010]_{\rm o}}$ sites [defined in Fig.~\ref{fig:AsRotationSP}(a)] are inequivalent, while they are equivalent for $H\parallel[110]_{\rm o}$ and $[1\bar{1}0]_{\rm o}$. In Ba(Fe$_{1-x}$Co$_{x}$)$_{2}$As$_{2}$, NNs to Co dopants produce similar peaks~\cite{JulienEPL,LaplacePRB,NingPRL}. The angle dependence of the resonance frequency of the As$_{1}$ center peak is reproduced with the parameters $|\nu_{z}|=5$ MHz, $\eta=0.24$, and $\delta=30^{\circ}$, where $\delta$ is the tilt angle of the principal axis ($z$-axis) of the EFG from the crystalline $c$-axis [Fig.~\ref{fig:AsRotationSP}(b)] to [100]$_{\rm o}$.
The Knight shift of As$_{1}$ is assumed to be the same as that of As$_{0}$. %since the identical $T$ dependence of the nuclear spin-lattice relaxation rate at both As and P sites \cite{NakaiPRB81} suggests that P-dopants do not significantly disturb the electronic states at the NN As sites.
These parameters enable us to estimate the positions of the As$_1$ satellite peaks [Fig.~\ref{fig:AsRotationSP}(c)].
As seen in the satellite peaks for $H\parallel[100]_{\rm o}$, weak and broad tails are observed. 
We consider that the broad tails arise from the As$_0$ site next to As$_1$ in the doped P and As$_1$ directions, since these tails have the same angle dependence as the As$_{1}$ site.
It is clear from Fig.~\ref{fig:AsRotationSP}(c) that the central part of each satellite signal essentially arises from the As$_{0}$ sites where the effect from the P-dopant is small.

Below, we discuss the electronic state of the As$_{0}$ sites by analyzing two quantities: (i) the absolute value of their quadrupole frequency $|\nu|$, directly determined from half of the separation between the high- and low-frequency satellites $|\nu(\phi)|=|f_{\frac{3}{2}\leftrightarrow\frac{1}{2}}-f_{-\frac{1}{2}\leftrightarrow-\frac{3}{2}}|/2=|\nu_{c}|(1+\eta\cos2\phi)/2$, (ii) the distribution  of quadrupole frequencies ${\Delta\nu}/|\nu|$, where $\Delta\nu$ is the full-width at half-maximum obtained by the Gaussian fit of the As$_{0}$ satellite peak.

The central observation of our study is shown in Figs.~\ref{fig:TemperatureSP}(a) and \ref{fig:TemperatureSP}(b). 
While the values of $|\nu|$ are almost identical for $H\parallel[110]_{\rm o}$ and for $H\parallel[100]_{\rm o}$, the values of $\Delta\nu/|\nu|$ for these two orientations are different and, most importantly, this difference increases on cooling.
The difference in the linewidth is independent of the analysis procedure. 
The contrasting behavior between $H\parallel[110]_{\rm o}$ and $H\parallel[100]_{\rm o}$ is also observed in the linewidth of the $^{75}$As-NMR center-line arising from the As$_0$ site measured in 6 T, as shown in Fig.~\ref{fig:Fig3C}.
We investigated the origin of the linewidth difference from the field dependence of the linewidth, and confirmed that the central linewidth is not determined by the hyperfine interaction, but mainly by the EFG. 
Whereas the linewidth decreases for $H\parallel[110]_{\rm o}$, that the width for $H\parallel[100]_{\rm o}$ is anomalously large and that it grows with decreasing $T$ can only be explained if there are two unresolved sites for $H\parallel[100]_{\rm o}$ having identical EFG tensor components but a nonzero $\eta$ and in-plane principal axes perpendicular to each other, although no clear splitting was observed in the satellite spectra at 130 K. 
The experimental results can be understood by considering the presence of perpendicular domains, in each of which the in-plane fourfold symmetry of the EFG is broken. 
Note that the central and satellite lines were found to have identical widths for $H\parallel[110]_{\rm o}$ and for $[1\bar{1}0]_{\rm o}$ (Fig.~\ref{fig:AsRotationSP} (b) and \ref{fig:AsRotationSP} (c)). 
%%%%%%%%%%%%%%%%%%%%%%%%%%%%%%%%%%%%%%% Fig.4 %%%%%%%%%%%%%%%%%%%%%%%%%%%%%%%%%%%%%%%%%%%%
\begin{figure}[t]
\begin{center}
\includegraphics[width=\hsize,clip]{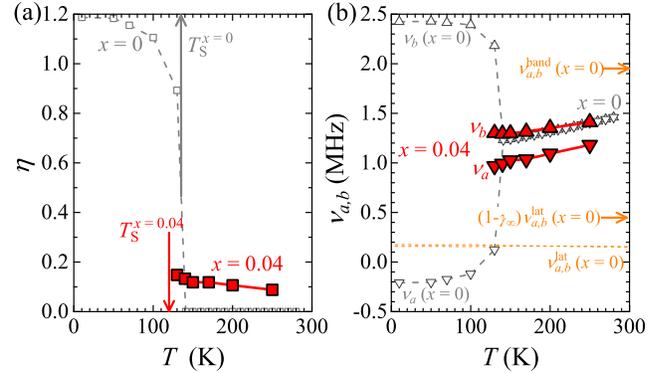}
\caption{(Color online) (a) $T$ dependence of in-plane EFG anisotropy parameter $\eta$ for $x=0.04$ (this work) and $x=0$ (from Ref.~45). For $x=0.04$, $\eta$ is derived from the angular dependence of the satellite linewidth with respect to the field. For $x=0$, however, it is deduced from the satellite splitting owing to twinned orthorhombic domains~\cite{KitagawaJPSJ77}. If domains are present above $T_{\rm S}$, this method does not allow the detection of $\eta\neq0$ locally.
(b) $T$ dependence of $\nu_{a,b}$ for $x=0.04$ together with the $x=0$ data from Ref.~45. $\nu_{a,b}^{\rm band}$ is the quadrupole frequency estimated from band-structure calculations. $\nu_{a,b}^{\rm lat}$ is the value calculated using a point charge model and the lattice parameters from Refs.~50 and 51. $(1-\gamma_{\infty})\nu_{a,b}^{\rm lat}$ for $x = 0$ is derived from the $\nu$ vs. $K^{\rm spin}$ plot (see Supplemental Material)\cite{SupplementalMaterial}. }
\label{fig:eta}
\end{center}
\end{figure}
%%%%%%%%%%%%%%%%%%%%%%%%%%%%%%%%%%%%%%%%%%%%%%%%%%%%%%%%%%%%%%%%%%%%%%%%%%%%%%%%%%

%%\section{Analysis}
Thus, to estimate the EFG anisotropy parameter $\eta$, we performed new fits to the sharp part of the satellite peaks ({\it i.e.}, excluding the tails since the P-dopant effect is observed) for $H\parallel[100]_{\rm o}$ with two Gaussian functions, each having a linewidth fixed to the value measured for $[110]_{\rm o}$ [Figs.~\ref{fig:TemperatureSP}(c), \ref{fig:TemperatureSP}(d)].
The flat-top shape of the spectra is well reproduced by the two- Gaussian peaks.
The corresponding two sites are named As$_{0}^{a}$ and As$_{0}^{b}$ and they are defined by their respective quadrupole frequencies $|\nu_{a}|<|\nu_{b}|$ for $H\parallel[100]_{\rm o}$.
It is important to realize that the frequency difference between the satellite peaks of As$_{0}^{a}$ and As$_{0}^{b}$ actually corresponds to $|\nu_{a}-\nu_{b}|$ for perpendicular domains in which the $a$- and $b$- axes are rotated by 90$^\circ$.
By dividing this value by $|\nu_{c}|=|\nu_{a}+\nu_{b}|=2|\nu_{[110]_{\rm o}}|$, $\eta$ can be estimated.
As shown in Fig.~\ref{fig:eta}(a), $\eta$ is already nonzero at 250~K but it increases on cooling. 

In general, the EFG has two origins: $\nu_{\alpha}=\nu^{\rm on}_{\alpha}+(1-\gamma_{\infty})\nu^{\rm lat}_{\alpha}$, where $\nu^{\rm on}_{\alpha}$ is the contribution from the on-site charge density and $\nu^{\rm lat}_{\alpha}$ is the contribution from the surrounding charged ions in the lattice, multiplied by the antishielding factor $(1-\gamma_{\infty})$~\cite{SternheimerPR}. 
Therefore, in-plane anisotropy ($\eta \neq 0$) can occur because of nematic order, ({\it i.e.}, electronic states locally break the fourfold symmetry) and/or the locally orthorhombic lattice symmetry (while still globally tetragonal).
%%%%%%%%%%%%%%%%%%%%%%%%%%%% Fig. 5%%%%%%%%%%%%%%%%%%%%%%%%%%%%%%%
\begin{figure}[t]
\begin{center}
\includegraphics[width=\hsize,clip]{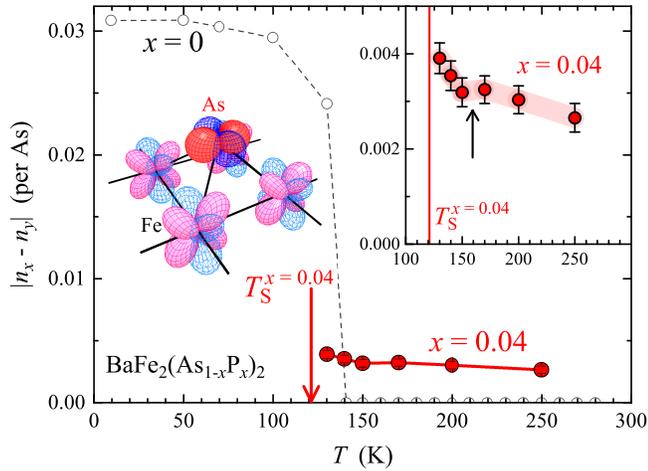}
\caption{(Color online) $T$ dependence of the orbital polarization $|n_{x}-n_{y}|$ for $x=0$ and 0.04 estimated by $\nu_{\alpha}^{\rm on}$ and Eq.~(\ref{eq:polarization}).
The right inset shows $x=0.04$ data on an enlarged scale. The black arrow shows a crossover in the $T$ dependence, coinciding with observations of a nematic state in bulk measurements \cite{KasaharaNATURE486}.
The left inset is a schematic description of Fe-$3d$ and As-$4p$ orbital orderings.
}
\label{fig:occupation}
\end{center}
\end{figure}
%%%%%%%%%%%%%%%%%%%%%%%%%%%%%%%%%%%%%%%%%%%%%%%%%%%%%%%%%%%%%%%%%%%

%in order to check the consistency and determine the sign of experimentally obtained $\nu_{a,b}$ by comparing theoretical $\nu_{a,b}$ for $x=0$

In order to obtain information on the on-site wave function, we examine the relative weights of these two contributions.
We performed band-structure calculations for $x=0$ within the local density approximation (LDA) for non-spin-polarized BaFe$_2$As$_2$\cite{BlahaWIEN2K} and using the lattice parameters reported previously~\cite{RotterPRB}. 
Since the obtained $\nu^{\rm band}_{a,b}=+1.95$~MHz at 300~K is similar to the experimental value $|\nu_{a,b}|=1.5$~MHz~\cite{KitagawaJPSJ77}, it is found that the on-site contribution is dominant, and the sign of $\nu_{a,b}$ is chosen to be positive in Fig.~\ref{fig:eta}(b) for both $x=0$ and 0.04 in the tetragonal phase. 
On the basis of theoretical~\cite{TakahashiJPSJ44} and experimental~\cite{MajumderARXIV} grounds, we estimate the $T$-independent lattice contribution $(1-\gamma_{\infty})\nu^{\rm lat}_{a,b}=0.45$~MHz from the extrapolation of the linear relationship between $\nu$ and the spin part of the Knight shift $K^{\rm spin}$ (with $T$ as an implicit parameter) to $K^{\rm spin} \rightarrow 0$ for the parent compound $x=0$ (see Supplemental Material)\cite{SupplementalMaterial}.  
This contribution thus accounts for less than half of the total $\nu_{a,b}$ in both $x=0$ and 0.04.
In addition, the observed nonzero $\eta$ value above $T_{\rm S}$ is much larger than $\eta$ estimated from the point-charge model using the lattice distortion reported in Ref. 22. 
Taking these results into account, the main origin of the in-plane anisotropy is due not to local orthorhombic distortion, but to the twofold symmetry of the electronic states.

$\nu_{\alpha}^{\rm on}$ is then deduced by subtracting $(1-\gamma_{\infty})\nu^{\rm lat}_{\alpha}$ from the experimental values of $\nu_{\alpha}$ for $x=0$ and 0.04.
By defining $i=(x,y,z)$ with $4p_{x}$ and $4p_{y}$ orbitals of As directed toward As NN, $\nu_{\alpha}^{\rm on}$ can be expressed in terms of the occupation number $n_{i}$ for each As-$4p_{i}$ orbital~\cite{HanzawaJPSJ59,ZhengJPSJ64}, and the $n_i$ and the orbital polarization corresponding to an occupation difference between As-$4p_{x}$ and $4p_{y}$ are estimated from   
\begin{equation} 
\nu_{(a,b,c)}^{\rm on}\!=\!\nu_{0}\!\left(\!n_{(x,y,z)}\!-\!\frac{n_{(y,z,x)}+n_{(z,x,y)}}{2}\!\right) 
\label{eq:occupation}
\end{equation}
and
\begin{equation} 
|n_{x}-n_{y}|=\frac{2}{3\nu_{0}}|\nu_{a}^{\rm on}-\nu_{b}^{\rm on}|,
\label{eq:polarization}
\end{equation}
where $\!\nu_{0}\!=\!\frac{3}{20}\frac{e^{2}Q}{h}\frac{1}{4\pi \epsilon_{0}}\frac{4}{5}\langle r^{-3}\rangle_{4p}\!=\!56.90 \ \rm{MHz}$ with $\langle r^{-3}\rangle_{4p}=6.958$ a.u.
Note that this simple modeling neglects any spatial modulation of the amplitude of the orbital polarization and the possibility that it is zero for some of the nuclei (in which case we should fit with three lines in an all-or-nothing model). 
Therefore, the value of $|n_x-n_y|$ should only be considered as an order-of-magnitude estimation. 
In particular, $|n_x-n_y|$ could be overestimated at high temperatures and thus it would extrapolate to zero at a significantly lower temperature than what the inset of Fig.\;5 suggests. 
%Note that in doing so, we attribute all of the $T$-dependent broadening for $H\parallel[100]_{\rm o}$ to the splitting, thus to the magnitude of the orbital polarization $|n_{x}-n_{y}|$.
%However, part of the broadening must be accounted for by the $T$ dependence of the correlation length. Taking this effect into account is clearly beyond the scope of this paper.
%Hence, the values of $|n_{x}-n_{y}|$ reported in Fig.~\ref{fig:occupation}  should be considered as lower bounds.

%\section{Discussion}
$|n_{x}-n_{y}|$ for $x=0.04$ is much smaller than in the orthorhombic phase of $x=0$, but it clearly increases with decreasing $T$, especially below $\sim160$~K. 
This is clearly observed in the linewidth of the central transition shown in Fig.\;3.    
This characteristic temperature of $\sim160$~K is consistent with the temperature at which twofold symmetry appears in the previous magnetic torque, lattice parameters, and Fe-$3d$ orbital polarization~\cite{KasaharaNATURE486,ShimojimaARXIV}. 
However, more work is needed to determine whether there is a continuous change in the magnitude and length scale of the orbital polarization (our data can be fit to a Curie Weiss law) or whether there is a sharp change at $\sim 160$ K. 

While quantitative evaluation of the polarization of the Fe-$3d$ orbitals has been reported in a few experiments for the AFM ordered state~\cite{JensenPRB,ZhangPRB85,YiPNAS}, no such measurement has been reported for the As-$4p$ orbitals above $T_{\rm S}$, to the best of our knowledge. 
Furthermore, the low time scale of NMR (set by the inverse linewidth of $\sim0.5$ MHz) implies that the observed orbital ordering is static. 
The imbalance between $4p_{x}$ and $4p_{y}$ occupations most likely results from the ordering within the Fe-$3d$ orbitals, that is, the degeneracy of Fe-$3d_{zx}$ and $3d_{yz}$ orbitals is lifted. 
Indeed, LDA calculations revealed that a reduction of the Fe-$3d_{yz}$ partial density of states (PDOS) is accompanied by a decrease in the As-$4p_{y}$ PDOS in the AFM ordered state for $x=0$~\cite{ShimojimaPRL}. 
This suggests that orbital ordering above $T_{\rm S}$ for BaFe$_{2}$(As$_{0.96}$P$_{0.04}$)$_{2}$ produces $n_{x}<n_{y}$ for the As-$4p$ orbitals.
This in-plane orbital anisotropy must participate in lifting the degeneracy of stripe spin correlations between $[100]_{\rm o}$ and $[010]_{\rm o}$ directions. 
A similar relationship between the lattice symmetry breaking and the enhancement of spin fluctuations has been observed in LaFeAsO~\cite{NakaiPRB85,FuPRL}. 
%In any event, our point here is that static orbital ordering is present and it must thus be taken into account as an important ingredient of the physics of FeAs planes.

%Furthermore, in NaFeAs, Zhang {\it et al.} (PRB 2012) find an orbital reconstruction in the nematic state affecting primarily $d_{yz}$ and $d_{xy}$ bands but not $d_{xz}$.}

An important and unanticipated aspect of our data is that $\eta$ at the As$_{0}$ sites, albeit small, is nonzero even at the highest measured temperature of 250~K, that is, well above the claimed nematic transition at $T^{*}\sim$~160~K~\cite{KasaharaNATURE486}.  
We suspect that orbital ordering would be entirely fluctuating at high temperatures in the absence of disorder but it is made static owing to pinning by defects (most likely P dopants) and/or microstrain induced by the dopants \cite{InosovPRB}. 
Such a phenomenon is analogous to the pinning of charge-density-wave (CDW) fluctuations above the CDW phase transition\cite{BerthierJPCSSP,WuPREPRINT} .
%or to the pinning of AFM fluctuations around impurities in high-$T_{\rm c}$ cuprates and low-dimensional antiferromagnets~\cite{JulienPRL,AlloulRMP,TedoldiPRL}. 
In iron-based superconductors, there is actually theoretical~\cite{InouePRB,GastiasoroARXIV} and experimental evidence from scanning tunneling spectroscopy~\cite{RosenthalARXIV,YangPRB86,ChuangSCIENCE,AllanNATUREPHYS} and transport measurements~\cite{IshidaPRL} that defects play a role in the nematic response. 
We stress that intrinsic spin and/or orbital nematic correlations are not produced by defects but they can be made visible by them. 
We expect this to be particularly true in the absence of any aligning field, {\it i.e.}, in the unstressed tetragonal phase. 
In a simplified picture, static, but short-range correlated, orbital order is nucleated around P sites. 
%Its magnitude, proportional to the amplitude of the intrinsic nematic susceptibility of the pure system, and its spatial extent, set by the nematic correlation length of the pure system, both grow on cooling. 
%Here, given the mean P-P distance as small as five lattice spacings for 4\% doping, short-range orbital order may concern essentially all of the As sites and not only the P nearest neighbors, over a wide temperature range.

%\section{Summary}
To summarize, our NMR work revealed unequal populations of the As $4p_{x}$ and $4p_{y}$ orbitals owing to Fe-$3d$ orbital ordering in the (unstressed) tetragonal phase of BaFe$_{2}$(As$_{0.96}$P$_{0.04}$)$_{2}$. 
The data reveal that these orbital-nematic correlations are involved in the appearance of a nematic state detected below $\sim$160~K by other probes~\cite{KasaharaNATURE486,ShimojimaARXIV}. 
Furthermore, the magnitude of the orbital polarization, its static nature, and its persistence at temperatures well above 160~K place strong constraints on microscopic models of nematic order in this class of superconductors.

%\section*{Acknowledgements}
The authors thank R. Fernandes, Y. Gallais, H. Kontani, S. Onari, Y. Ohno, Y. Kobayashi, and M. Sato for valuable discussion, and O. Leynaud for X-ray measurements.
This work was supported by Scientific Research grant from JSPS and by the French ANR Supratetrafer (ANR-09-BLAN-0211), Kyoto Univ. LTM center, a Grant-in-Aid for the Global COE Program ``The Next Generation of Physics, Spun from Universality and Emergence'' from MEXT of Japan, and Grants-in-aid for Scientific Research from the Japan Society for the Promotion of Science (JSPS), KAKENHI (S and A) (Nos. 20224008 and 23244075). 
One of the authors (T.I.) is financially supported by a JSPS Research Fellowship.

%\bibliographystyle{jpsj.bst}
%\bibliography{SingleCrystal}

\begin{figure}[b]
\begin{center}
\includegraphics[width=70mm,clip]{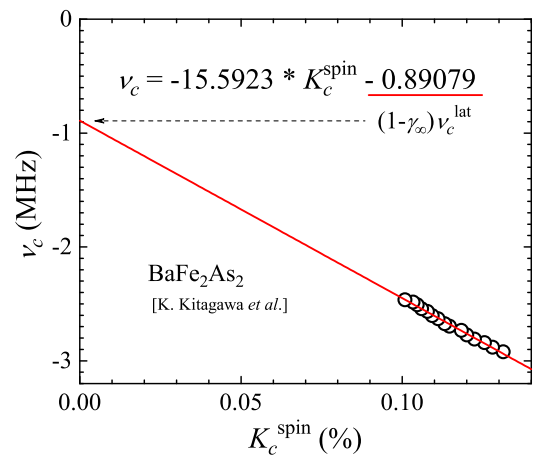}
\caption{(Color online) The linear relationship between $\nu_{c}$ and $K^{\rm spin}_{c}$ as an implicit function of $T$. A straight red line indicates the linear fitting of the data (open circles) extrapolated down to $K^{\rm spin}_{c}\rightarrow0$.
The fitting equation with obtained parameters is also shown in the figure. The underlined value indicates the intercept, i.e., $(1-\gamma_{\infty})\nu_{c}^{\rm lat}$.
}
\label{fig:K-Nuplot}
\end{center}
\end{figure}

\begin{center}
\large\bf{Supplemental Material for \\
``Emergent of Orbital Nematicity in the Tetragonal Phase of BaFe$_2$(As$_{1-x}$P$_x$)$_2$''
}
\end{center}
Majumder {\it et al}. reported the linear relationship between $K^{\rm spin}$ and $\nu$ with an implicit parameter $T>T_{\rm S}$ in iron pnictides \cite{MajumderARXIV}.
It was indicated from theoretical point of view that the temperature dependence of $\nu$ can be influenced by the spin susceptibility in the presence of the mode-mode coupling between charge and spin density fluctuations in itinerant magnets \cite{TakahashiJPSJ44}.
Such linear relationships are actually observed in the various itinerant magnet systems like MnSi \cite{YasuokaJPSJ44}, PrCoAsO \cite{MajumderARXIV}, NaCo$_2$O$_4$ \cite{RayPRB}, LiFeAs \cite{BaekEPJ}, and BaFe$_2$As$_2$ \cite{KitagawaJPSJ77}. 
Since the existence of the mode-mode coupling between these two channels seems to be reasonable, the contribution of $(1-\gamma_{\infty})\nu_{\alpha}^{\rm lat}$ for $x=0$ can be extracted by extrapolating the linear portion of $\nu_{\alpha}$ versus $K^{\rm spin}_{\alpha}$ plot to $K^{\rm spin}_{\alpha}\rightarrow0$. 
For the estimation, we used $\nu_{c}$ and $K^{\rm spin}_{c}$ adopted from Ref. \cite{KitagawaJPSJ77} to derive $(1-\gamma_{\infty})\nu_{c}^{\rm lat}$ (Fig. \ref{fig:K-Nuplot}). 
Then, $(1-\gamma_{\infty})\nu_{a,b}^{\rm lat}$, which is the value used in Fig. 4(b), is derived by multiplying $-1/2$ to $(1-\gamma_{\infty})\nu_{c}^{\rm lat}$ by following the relationship $\nu_{a,b}^{\rm lat}=-(1/2)\times\nu_{c}^{\rm lat}$ from Laplace's equation $\nu_{a}^{\rm lat}+\nu_{b}^{\rm lat}+\nu_{c}^{\rm lat}=0$ and $\nu_{a}^{\rm lat}=\nu_{b}^{\rm lat}$ above $T_{\rm S}$.
Finally, we obtain $(1-\gamma_{\infty})\nu_{a,b}^{\rm lat}$ to be 0.45 MHz.

\bibliographystyle{jpsj.bst}
\bibliography{SingleCrystal}

\end{document}